\documentclass[fleqn,usenatbib]{mnras}

\usepackage{newtxtext,newtxmath}

\usepackage[T1]{fontenc}

\DeclareRobustCommand{\VAN}[3]{#2}
\let\VANthebibliography\thebibliography
\def\thebibliography{\DeclareRobustCommand{\VAN}[3]{##3}\VANthebibliography}

\usepackage{graphicx}	
\usepackage{amsmath}	
\usepackage[acronym]{glossaries}
\usepackage{xspace}
\usepackage{listings}
\usepackage[dvipsnames]{xcolor}
\usepackage[normalem]{ulem}

\usepackage[nameinlink,noabbrev]{cleveref}
\crefname{equation}{Eq.}{Eqs.}
\crefname{section}{Sect.}{Sects.}
\crefname{figure}{Fig.}{Figs.}
\crefname{table}{Table}{Tables}
\crefname{appendix}{Appendix}{Appendices}
\Crefname{figure}{Figure}{Figures}
\Crefname{equation}{Equation}{Equations}
\Crefname{section}{Section}{Sections}
\Crefname{table}{Table}{Tables}

\definecolor{codegreen}{rgb}{0,0.6,0}
\definecolor{codegray}{rgb}{0.5,0.5,0.5}
\definecolor{codepurple}{rgb}{0.58,0,0.82}
\definecolor{backcolour}{rgb}{0.93,0.93,0.90}
\definecolor{backcolour}{rgb}{0.95,0.95,0.92}

\lstdefinestyle{mystyle}{
    frame=single,
    backgroundcolor=\color{backcolour},   
    commentstyle=\color{codegreen},
    keywordstyle=\color{magenta},
    numberstyle=\tiny\color{codegray},
    stringstyle=\color{codepurple},
    basicstyle=\ttfamily\footnotesize,
    breakatwhitespace=false,         
    breaklines=true,                 
    captionpos=b,                    
    keepspaces=true,                 
    numbers=none,                    
    numbersep=5pt,                  
    showspaces=false,                
    showstringspaces=false,
    showtabs=false,                  
    tabsize=2
}

\newcommand{\code}[1]{\texttt{#1}}
\newcommand{\Abacus}{\textsc{AbacusSummit}\ }
\newcommand{\Gpch}{\ h^{-1}\text{Gpc}}
\newcommand{\Mpch}{\ h^{-1}\text{Mpc}}
\newcommand{\hhhMpc}{\ h^3\,\text{Mpc}^{-3}}
\newcommand{\glscite}[2]{%
  \ifglsused{#1}
  {%
    \glsentryshort{#1}%
  }{%
    \glsentrylong{#1}\ (\glsentryshort{#1}; \citealt{#2})%
    \glsunset{#1}%
  }%
  \xspace
}

\glsdisablehyper
\newacronym{VSF}{VSF}{void size function}
\newacronym{RSD}{RSD}{redshift space distortion}
\newacronym{AP}{AP}{Alcock-Paczy\'nski}
\newacronym{BOSS}{BOSS}{Baryon Oscillation Spectroscopic Survey}
\newacronym{DESI}{DESI}{Dark Energy Spectroscopic Instrument}
\newacronym{LRG}{LRG}{luminous red galaxy}
\newacronym{BAO}{BAO}{Baryon Acoustic Oscillation}
\newacronym{STH}{STH}{spherical top-hat}
\newacronym{FFT}{FFT}{Fast Fourier Transform}
\newacronym{HOD}{HOD}{halo occupation distribution}
\newacronym{TSC}{TSC}{Triangular Shaped Cloud}
\newacronym{NGP}{NGP}{Nearest Grid Point}
\newacronym{SGC}{SGC}{South Galactic Cap}

\title[Void-finding with VERSUS]{VERSUS: An excursion-set-inspired void-finder for the Stage-IV era}

\author[N. Findlay and S. Nadathur]{
Nathan Findlay$^{1}$\thanks{E-mail: nathan.findlay@port.ac.uk}
and Seshadri Nadathur$^{1}$
\\
$^{1}$Institute of Cosmology and Gravitation, University of Portsmouth, Burnaby Road, Portsmouth PO1 3FX, UK
}

\date{Accepted 2026 July 22. Received 2026 June 29; in original form 2026 May 5}

\pubyear{\the\year{}}

\begin{document}
\label{firstpage}
\pagerange{\pageref{firstpage}--\pageref{lastpage}}
\maketitle

\begin{abstract}
We present \code{VERSUS}, a publicly available, fast void-finding algorithm designed to identify spherical underdensities in the density field that can be accurately described by excursion set predictions of the void size function. We validate the algorithm against both a synthetic distribution of particles designed to trace a known input void population, and mock galaxy sample built from a $(2\Gpch)^3$ \Abacus simulation populated with a realistic galaxy-halo connection, including systematic effects designed to mimic real survey data. In all cases, \code{VERSUS} demonstrates excellent performance, achieving strong agreement with theoretical predictions for the void size function across the range $25 < R \,[\Mpch] < 61$ without requiring any post-processing of the void catalogue. The code is user-friendly, modular, and readily applicable to observational survey data. Its computational efficiency further enables the use of simulation-based modelling approaches, facilitating robust and consistent cosmic void analyses with Stage-IV surveys.

\end{abstract}

\begin{keywords}
cosmology: large-scale structure of Universe -- cosmology:
observations -- methods: data analysis -- gravitation -- software: data analysis
\end{keywords}



\section{Introduction}

While dark matter nodes and filaments dominate the mass contribution in the Universe, it is the lowest density regions that occupy the majority of the volume \citep{Sheth2004}. Cosmic voids present a unique probe of non-Gaussian information in the galaxy distribution. By shining a light on the dynamics of the large-scale structure in underdense environments, voids gain sensitivity to diffuse components such as dark energy and neutrinos, and to environment-dependent effects such as modified gravity and galaxy evolution, making them an advantageous test of fundamental physics \citep{Pisani2019, Contarini2026}.

The distribution of void sizes, typically referred to as the \gls{VSF}, is directly sensitive to the cosmological model. Changes to the matter content \citep{Contarini2024} or the dark energy equation of state \citep{Pisani2015} lead to a different abundance of voids present in the matter distribution. Theoretically, a simple model for the matter \gls{VSF} can be derived from the statistical properties of the linear density field \citep{Sheth2004, Jennings2013}. However, this idealised case is difficult to match to real observed voids due to both tracer bias and algorithmic distinctions \citep{Furlanetto&Piran2006, Nadathur2015a, Nadathur2015b}. The definition of a void---and by extension the process for finding them---is not uniquely specified. The only necessary condition is that the region associated with a void must exist as an underdensity relative to the mean. The void centre and boundary definitions, however, are a consequence of the choice of detection algorithm. A huge variety of void finding algorithms exist on the market with no one-size-fits-all solution \citep{Colberg2008}. These algorithms, at minimum, provide a catalogue of void centre positions and corresponding void sizes with which cosmological analysis can be performed.

Topological void-finders, such as those based on the watershed algorithm, generally find rather large voids, since the boundaries of such voids are defined so as to include higher-density regions at their outskirts. Such topological void-finders are therefore often paired with a method for catalogue cleaning in post-processing to obtain voids that have properties closer to those assumed in theoretical models for the \gls{VSF}. The watershed-based algorithm \code{VIDE}\footnote{\url{https://bitbucket.org/cosmicvoids/vide_public/wiki/Home}} \citep{Sutter2015} is commonly used in studies of the \gls{VSF}, as described below. Using a tessellation of the tracer distribution with Voronoi cells in order to define local estimates of the density field, \code{VIDE} treats all density minima as potential void sites and defines the void volume from an ensemble of cells which are bounded by a saddle point region in density. Naturally, this results in a catalogue of non-spherical voids spanning a range of depths, with sizes set by a density turnover at their boundaries. Importantly, the algorithm is volume-filling, so all high-density clusters, filaments and walls are by construction included within the volume assigned to some voids of the resultant catalogue.

However, theoretical models of the \gls{VSF} are typically based on the excursion set formalism, in which hierarchical structure formation is captured by a stochastic random walk of the smoothed linear density field and voids are identified as regions lying below some minimum density threshold. Within this framework, void abundances and sizes are predicted from their integrated density contrast. To obtain a cleaner sample with radii more consistent with theoretical expectations, it is therefore necessary to prune the outputs of topological void-finders and apply density-based resizing \citep{Ronconi2017, Ronconi2019} as a further post-processing step. An improved version of the cleaning method that does not enforce strict sphericity of the resized void has also recently been proposed by \cite{Verza2025}. 

Alternative non-topological algorithms, aimed at producing void catalogues that have a better direct match to theoretical predictions without requiring this cleaning, have also been explored \citep{Ruiz2015, Paz2023, Ruiz2026}. However, most mature \gls{VSF} analyses (e.g. \citealt{Contarini2019, Bayer2021, Contarini2021, Verza2023,Contarini2023, Contarini2022, Verza2025}) currently continue to use topological void-finding algorithms such as \code{VIDE} plus a post-processing cleaning, together with analytic modelling of the VSF that incorporates simple assumptions about tracer bias around voids, and the effects of \glscite{RSD}{Kaiser1987} and \glscite{AP}{Alcock1979} distortion on void size. Some recent approaches have also leveraged simulation-based forward models of the \gls{VSF} \citep{Thiele2024, Fraser2025, Salcedo2025, Lehman2026}: this approach does not require a match to the void properties assumed in the excursion set approach and so can in principle use the output of any void-finder without additional cleaning. However, simulation-based approaches are computationally demanding with current algorithms, which---as we will discuss below---are slow and scale poorly with the number of galaxies given the requirements of large modern Stage-IV surveys such as the \glscite{DESI}{Snowmass2013, DESI2016} and \emph{Euclid} \citep{Euclid2011}.

In this work, we present a novel void-finding algorithm to target underdense regions of the galaxy field using a threshold on the integrated density contrast. It is based on an initial step identifying spherical underdensities in the galaxy distribution satisfying a specified density threshold criterion, and subsequently allows for optimal merging of overlapping spheres of different sizes, producing a final void catalogue that is not limited to only spherical structures. The code is provided publicly as a modular \code{Python} package with a streamlined user interface and support for both cubic simulation boxes with periodic boundary conditions and complex survey geometries and selection functions. The underlying routines are written in \code{C} and \code{Cython} for computational speed. The algorithm is accurate enough to achieve strong agreement with theoretical predictions of the \gls{VSF} using high-fidelity galaxy mock simulations, robust enough to handle complex survey data from Stage-IV surveys, and fast enough to enable simulation-based modelling approaches. We demonstrate that the algorithm provides a direct and self-consistent match to the excursion set theory across a large range of scales, with no post-processing of the void catalogue---a first for void-finders applied to the galaxy distribution.

The paper is structured as follows. In \autoref{sec:vsf_theory}, the excursion set theory employed to model the \gls{VSF} is summarised. This section crucially introduces the motivation behind the \code{VERSUS} algorithm, which is discussed in \autoref{sec:void_finding}. We compare the algorithm against common topological void-finders, performing the validation initially on a simple toy model simulation and then extending to a high-fidelity Stage-IV galaxy mock. The simulated data is described in \autoref{sec:simulations} and the results of the validation are discussed in \autoref{sec:results}. Finally, the conclusions are given in \autoref{sec:conclusions}.

\section{Theoretical void size function}
\label{sec:vsf_theory}

In order to motivate an algorithm to measure the \gls{VSF}, it is important to first have a theoretical description. 

\subsection{Voids in the matter field}

The excursion set formalism predicts the abundance of objects crossing a critical density threshold as a consequence of hierarchical structure formation in an initially linear Gaussian field \citep{Press&Schecter1974, Peacock1990, Bond1991, Lacey1993, Mo&White1996, Sheth&Tormen2002}. It was originally developed to describe the collapse of virialised structures, predicting the mass function of dark matter halos (see \citealt{Zentner2007} for a review).

\cite{Sheth2004} later extended this framework to cosmic voids by introducing an additional condition to exclude voids embedded within collapsing regions. Subsequent refinements by \cite{Jennings2013} enforced a physically consistent volume fraction, ensuring that the total void volume does not exceed unity, leading to the widely used \emph{Vdn} model. We outline the derivation of this model below.

By following the random walk of the density field as it is incrementally smoothed as a function of scale $R_L$,\footnote{The subscript $L$ here denotes that the smoothing applied to the \emph{linear} field to avoid confusion when quantities are transformed to the non-linear regime.}
\begin{equation}
    \delta^\mathrm{sm}(\vec{x}, R_L) = \int \frac{d^3k}{(2\pi)^3} W(\vec{k}, R_L) \delta(\vec{k}) e^{-i \vec{k}\cdot\vec{x}} ,
\end{equation}
where $W(\vec{k}, R_L)$ is the top-hat filter in Fourier space, the probability of the field crossing any given density threshold can be determined. The two thresholds of interest for void formation are the linear formation threshold, $\Delta^L_v$, and the linear halo (or collapse) threshold, $\Delta^L_c$. Specifically of interest are random walks that cross $\Delta^L_v$ without having previously crossed $\Delta^L_c$. This `double barrier' approach ensures that voids embedded within collapsing overdensities (void-in-cloud) are disregarded. The variance of the smoothed density field,
\begin{equation}
    \sigma_m^2(R_L) = \int \frac{k^2 dk}{2\pi^2} P^\mathrm{lin}_m(\vec{k}) {\left|W(\vec{k}, R_L)\right|}^2 ,
\end{equation}
can simply be related to the linear matter power spectrum, $P^\mathrm{lin}_m(\vec{k})$. Although these quantities only strictly describe a true random walk when $W(\vec{k}, R_L)$ is a Fourier space top-hat filter, they are typically instead defined in terms of a configuration space top-hat filter in order to reconcile the interpretation with observed voids (for further discussion on this topic, see Section 3.1 of \citealt{Contarini2026}). With the configuration space definition, the smoothed density contrast $\delta^\mathrm{sm}(\vec{x}, R_L)$ can be interpreted as the integrated (or enclosed) density contrast, $\Delta(\vec{x}, R_L)$.

Following this we can derive the multiplicity function, an analytic expression for the fraction of random walks that cross $\Delta^L_v$ for the first time, having not previously crossed $\Delta^L_c$:
\begin{equation}
    \label{void_frac}
    f(\sigma_m) = 
    2\sum_{i=1}^{\infty} e^{-\frac{(i\pi x)^2}{2}} 
    i \pi x^2 \sin{(i\pi \mathcal{D})} \,
\end{equation}
with
\begin{align}
    \mathcal{D} = \frac{|\Delta^L_v|}{\Delta^L_c + |\Delta^L_v|},
    &&
    x = \frac{\mathcal{D}}{|\Delta^L_v|} \sigma_m .
\end{align}
The threshold values, derived in an Einstein-de Sitter model, are commonly set to $\Delta^L_v=-2.71$, corresponding to shell-crossing, and $\Delta^L_c = 1.69$, corresponding to complete collapse. The linear void threshold can be mapped to its non-linear counterpart, $\Delta_v$, using the fitting formula in \cite{Bernardeau1994}:
\begin{equation}
    (1 + \Delta_v) = {\left( 1 - \frac{\Delta^L_v}{\mathcal{C}} \right)}^{-\mathcal{C}} , 
\end{equation}
with $\mathcal{C}=1.594$, and thus $\Delta_v\simeq-0.8$. For derivations of the values adopted here, see the appendices of \cite{Jennings2013}.

This crossing fraction can be related to the number density of voids in a given volume,
\begin{equation}
    \frac{dn_v}{d\ln{R_L}}
    = \frac{f(\sigma_m)}{V(R_L)} \frac{d\ln{\sigma_m^{-1}}}{d\ln{R_L}} ,
\end{equation}
where $V(R_L)$ is the spherical void volume in linear theory. Non-linear evolution leads to further expansion of voids. In the case of isolated spherical voids, this evolution follows $R = R_L(1 + \Delta_v)^{-1/3}$, which is obtained from mass conservation of a fixed Lagrangian patch under spherical evolution. Finally, assuming the conservation of total void volume during the transition to the non-linear regime, we arrive at the \emph{Vdn} model:\footnote{In \cite{Sheth2004}, the total void number is conserved during the transition to the non-linear regime. In contrast, \cite{Jennings2013} proposed that voids merge with their neighbours, conserving total volume instead.}
\begin{equation}
    \label{eq:VSF}
    \frac{dn_v}{d\ln{R}}
    = \left. \frac{f(\sigma_m)}{V(R)} \frac{d\ln{\sigma_m^{-1}}}{d\ln{R_L}} \right|_{R_L = R (1+\Delta_v)^{1/3}} .
\end{equation}

With this, the distribution of spherical, non-overlapping underdensities embedded in the matter field is entirely specified by the linear matter power spectrum, $P^\mathrm{lin}_m(\vec{k})$, and two density threshold parameters, $\Delta_v$ and $\Delta^L_c$. Recent theoretical developments have explored the combination of the excursion set formalism with peak theory to broaden the range of scales on which the \gls{VSF} can be accurately modelled. In this framework, the excursion set calculation is restricted to a subset of Lagrangian density minima and employs a `moving barrier' prescription \citep{Sheth2001}, replacing the fixed void formation threshold with a scale-dependent parametrisation calibrated against simulations. This approach has shown promising agreement with matter-only simulations down to redshift $z=0$ \citep{Verza2024}.

\subsection{Voids in the tracer distribution}

In practice, voids are measured in the distribution of galaxies, which serve as biased tracers of the underlying matter distribution. Relating depressions found in the matter field, characterised by an integrated density contrast $\Delta_v^\mathrm{m}$, to those found in the galaxy distribution is therefore non-trivial. \cite{Furlanetto&Piran2006} proposed redefining the linear threshold $\Delta_v^L$ used in theoretical models, previously fixed by the shell-crossing condition, in terms of the observed enclosed galaxy density. This provides a natural connection between theoretical voids defined in the underlying matter distribution and those identified in a biased tracer population.

Building on this, \cite{Contarini2019} introduced a simple linear relation between the integrated tracer density contrast $\Delta_v^\mathrm{t}$ and theoretical matter threshold $\Delta_v^\mathrm{m}$:
\begin{equation}
    \label{eq:Delta_tr}
    \Delta_v^\mathrm{t} = b_v \Delta_v^\mathrm{m},
\end{equation}
where the void bias,
\begin{equation}
    b_v(b_1) = B_\mathrm{slope} b_1 + B_\mathrm{offset} ,
\end{equation}
is a linear function of the large-scale effective linear bias of the sample, $b_1$. The best-fit values of (or priors on) the free parameters, $B_\mathrm{slope}$ and $B_\mathrm{offset}$, were determined by matching void density profiles measured from halo catalogues to the dark matter distribution at the same point across various redshifts:
\begin{equation}
    b_v \equiv \left \langle
    \frac{\Delta^\mathrm{t}_{v}(r=R)}{\Delta^m_{v}(r=R)} \right\rangle ,
    \label{eq:bias}
\end{equation}
where $R$ is the spherical void radius determined from the tracer sample and the brackets denote an average across $R$ values in the sample. This idea was further extended from halos to galaxies in \cite{Contarini2022} where they are fit as free parameters of the model, and applied to the final \glscite{BOSS}{BOSS2013} data release in \cite{Contarini2023}. 

One can also take the approach of effectively treating $\Delta_v^\mathrm{t}$ as a free parameter to be fit to data (e.g. \citealt{Pisani2015}). While this means the model is no longer truly self-consistent, it provides a practical way to marginalise over the bias-dependency of \autoref{eq:Delta_tr} and absorb residual discrepancies that arise when the void-finding algorithm departs from the assumptions underlying the theoretical prediction.

Throughout this work, we utilise the implementation of the model described above as implemented in the cosmological software library \code{CosmoBolognaLib}\footnote{\url{https://github.com/federicomarulli/CosmoBolognaLib}} \citep{CBL2016}. The library also includes the void catalogue cleaning procedure introduced by \cite{Ronconi2017}, which we apply to the output of the topological void-finders to enable a like-to-like comparison with \code{VERSUS}. The void threshold value of $\Delta_v=-0.8$ is used consistently for the computation of the theoretical prediction, the \code{VERSUS} void-finding step, and the void cleaning algorithm.

\section{Void-finding with VERSUS}
\label{sec:void_finding}

\begin{figure*}
    \centering
    \includegraphics[width=\linewidth]{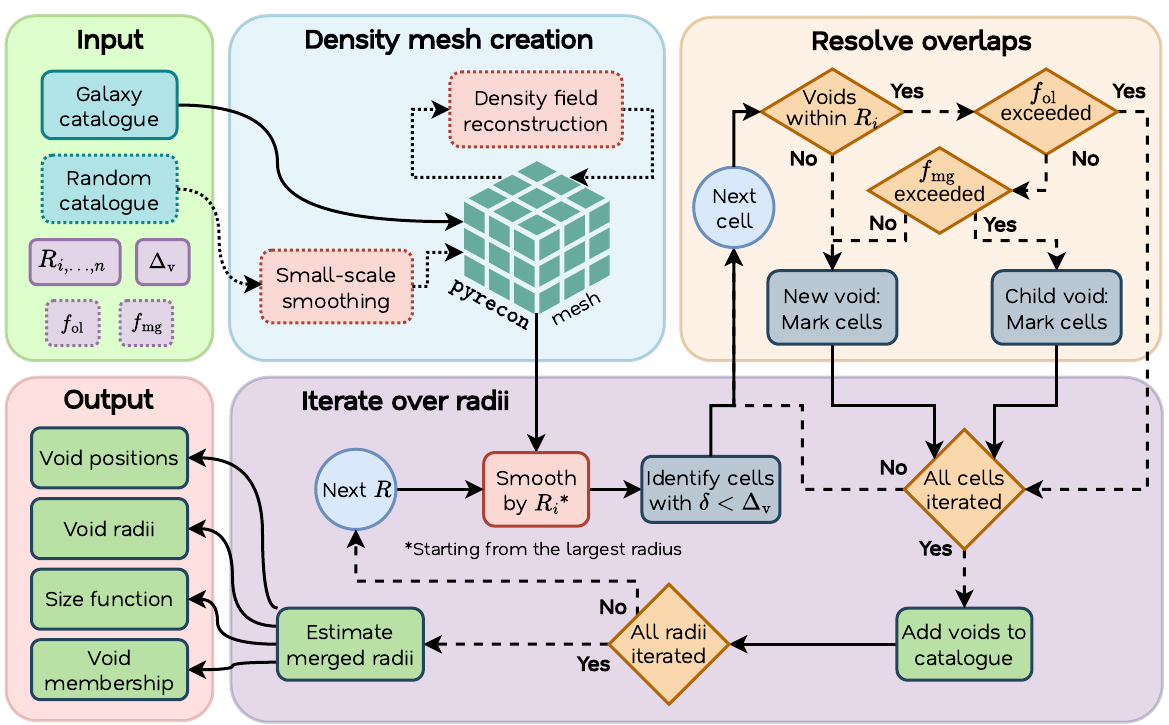}
    \caption{Flowchart illustrating the \code{VERSUS} void-finding algorithm. The code must be supplied with a set of input radii $R_{i,\dots, n}$ and an integrated density threshold $\Delta_v$, along with optional parameters controlling void overlap and merging, $f_\mathrm{ol}$ and $f_\mathrm{mg}$, respectively. Optional steps are shown with dotted boxes and arrows, decision points are indicated by dashed lines, and colours group related operations. All smoothing operations use the spherical top-hat filter, as defined in \autoref{eq:tophat}.}
    \label{fig:voidfinder}
\end{figure*}

In this section, we present the Void Extraction of Real-space Spherical UnderdensitieS (\code{VERSUS}\footnote{\url{https://github.com/ntbfin00/VERSUS}}) algorithm. Built on the foundations of the void-finder included in the \code{Pylians3}\footnote{\url{https://github.com/franciscovillaescusa/Pylians3}} package \citep{Pylians}, the code has been designed with the objective to match predictions of excursion-set-based models for the \gls{VSF}, and has added the functionality to enable consistent application in the presence of non-uniform selection functions. Primarily written in \code{Cython}, it is fast, modular and readily applicable to spectroscopic survey data.

\subsection{The algorithm}

The algorithm uses a series of iterative convolutions to rapidly extract spherically underdense regions from the tracer density field, using a simple merging criteria to consolidate regions of a given enclosed density. In its simplest form, it takes a catalogue of tracer positions, a set of input radii $R_i$, and a maximum integrated density threshold $\Delta_v$ as input. Optionally, settings such as the void overlap and merging fraction thresholds---$f_\mathrm{ol}$ and $f_\mathrm{mg}$, respectively---can be adjusted in order to modify the void-finding assumptions. Additionally, if the data has a complex survey footprint, a catalogue of unclustered `random' positions that sample the survey geometry and selection function can be supplied. The code returns void centre positions, void radii, a 3D array with void member cells marked, and the measured \gls{VSF}. \autoref{fig:voidfinder} provides a diagrammatic overview of the algorithm, which is described in greater detail below.

The code operates on a mesh, estimating the density field within cubic cells from a catalogue of 3D data positions, and, when supplied, a corresponding set of 3D random positions for realistic survey applications. All mesh-based computations are performed using the \code{pyrecon}\footnote{\url{https://github.com/cosmodesi/pyrecon}} package. We adopt a mesh cell size that is significantly smaller than the characteristic void sizes in a galaxy sample, ensuring that void radii are well resolved. By default, we use $4\Mpch$, which offers a practical balance between resolution and computational efficiency for void-finding in a $(2\Gpch)^3$ simulation. Nevertheless, a higher resolution can be advantageous where computationally feasible.

The random catalogue samples the survey geometry, selection function and potential observational systematic effects and is used to estimate both cell densities and the survey boundary when provided. These randoms are first assigned to the mesh using the \gls{TSC} interpolation scheme. Cells with a sparsity of randoms are treated as outside of the survey mask and flagged for later use. This procedure defines the survey boundary, which is subsequently reapplied after any density smoothing step, as well as the effective survey volume required for computing the \gls{VSF}. If the data are instead taken from a simulation box with periodic boundary conditions and uniform selection function, no randoms need be provided and this step is bypassed.

Once the survey boundary has been determined, data positions (and randoms if provided) are mapped to the mesh using \gls{NGP} assignment. Additionally, weights can be provided for both data and randoms if required. In order to mitigate noise in the random field, an initial smoothing, much smaller than the characteristic void size, is applied to the randoms before combining with the data to form the density field. All smoothings in \code{VERSUS} utilise \glspl{FFT} and the \gls{STH} kernel,
\begin{equation}
    \label{eq:tophat}
    W_\mathrm{STH}(k,R) = \frac{3}{(kR)^3} \left[ \sin(kR) - kR\cos(kR) \right] ,
\end{equation}
to efficiently filter the density field the in Fourier space.
The initial random smoothing scale $R_\mathrm{sm} = Sr_\mathrm{sep}$ is set as a fraction $S$ of the average random separation,
\begin{equation}
    \label{eq:r_sep}
    r_\mathrm{sep} = \Big(\frac{4 \pi n_r}{3}\Big)^{-\frac{1}{3}} ,
\end{equation}
where $n_r$ is the number density of random particles.
We find that choosing $S = 4$ is sufficient to obtain unbiased measurements of the \gls{VSF} when using a random catalogue with complex survey geometry (see \autoref{sec:results_performance}). In \cref{sec:sims_cutsky}, we adopt a random catalogue with a number density $50\times$ that of the galaxy tracers, which are constructed to match the \gls{DESI} \gls{LRG} sample \citep{Zhou2023}. As a result, $n_r \simeq 0.025 \hhhMpc$, which corresponds to a smoothing scale of $R_\mathrm{sm} = 8.5 \Mpch$, roughly twice the mesh cell size. After smoothing the random density field, $\rho_R$, it is combined with the data density field, $\rho_D$, to estimate the final overdensity field, 
\begin{equation}
    \delta = \frac{\rho_D - \rho_R}{\alpha \rho_R},
\end{equation}
where $\alpha=N_D/N_R$ is the ratio of the weighted number of data positions to random positions.

Although we do not explore this feature here, the \code{pyrecon} algorithm enables the approximate removal of \gls{RSD} effects. Density field reconstruction methods, developed for sharpening the \glscite{BAO}{Eisenstein2007, Padmanabhan2012} feature using the Zeldovich approximation \citep{Zeldovich1970}, have been shown to improve the accuracy of real-space void positions recovered from the galaxy distribution in redshift-space \citep{Nadathur2019a, Nadathur2019c}. In future work, we intend to explore the accuracy and information gain of this procedure.

Once the initial density mesh is generated, it is convolved with a series of \gls{STH} smoothings with scales equal to the selected input radii, representing the size bins within which we wish to identify voids. The input radii should be binned finely enough to adequately sample the \gls{VSF}. However, excessively fine binning can result in low bin counts and skew the measured \gls{VSF}, as uncertainties in void sizes may exceed the bin width (see the discussion in \autoref{sec:bias}). We set the default range to $R_v = [25, 61] \Mpch$ with a bin width of $\Delta R_v = 2 \Mpch$, as we have found this to be adequate for tracer densities of $n \sim 5 \times 10^{-4} \hhhMpc$ (see \autoref{sec:results_synthetic}). We adopt $R_v = 25 \Mpch$ as the minimum radius for this sample, as tests on the simulations in \autoref{sec:sims_synthetic} indicate that smaller radii can admit spurious voids given the sample’s number density. 
The maximum radius, $R_v = 61 \Mpch$, is chosen to include the largest voids expected for the galaxy sample we are investigating. While voids of this size are rare, the theoretical model described in \autoref{sec:vsf_theory} predicts $\mathcal{O}(10)$ voids in a $(2\Gpch)^3$ volume for a biased galaxy sample. Topological algorithms, however, have been shown to identify much larger ($> 100 \Mpch$) voids \citep{Nadathur2015a}, which are difficult to reconcile with theoretical predictions without post-processing of the resulting void catalogue.

Starting with the largest input radius $R_i$, an \gls{FFT} is applied to the mesh and then the Fourier amplitude in each cell is multiplied by the \gls{STH} kernel. The field is then inverse Fourier transformed to return the field smoothed on the scale of interest---directly mirroring the excursion set theory. If randoms were provided, the survey mask is now reimposed, setting $\delta=0$ in cells outside the survey. 

Cells with a smoothed overdensity $\delta^\mathrm{sm}$ less than the input threshold value $\Delta_v$ are then targeted as candidate void centres. For each cell below the threshold, starting from the least dense, the degree of overlap with previously detected voids is determined. Depending on the candidate radius, we quantify the fractional overlap, $\hat{f}_\mathrm{ol}$ , either by determining distances to other voids, or by counting the cells within its radius that have already been assigned to other voids. If $\hat{f}_\mathrm{ol}>f_\mathrm{ol}$, then the candidate is rejected. By default, overlaps up to the void centres are permitted; however, this setting can be easily adjusted by the user to allow only a fractional overlap or to enforce a strict no-overlap condition. 

A merging criterion---implemented by counting the number of cells shared between a candidate and existing voids---controls the extent to which voids may overlap while still being treated as distinct objects. 
If the region of radius $R_i$ centred on the candidate cell does not exceed the overlap threshold for merging, i.e. $\hat{f}_\mathrm{ol}<f_\mathrm{mg}$, the candidate is identified as a new void; otherwise, it is assigned as a child void of one of the overlapping parent voids. All cells within a distance of $R_i$ of the central candidate cell are then assigned to that void, and the algorithm proceeds to the next least-dense candidate cell.

\begin{figure}
    \centering
    \includegraphics[width=\linewidth]{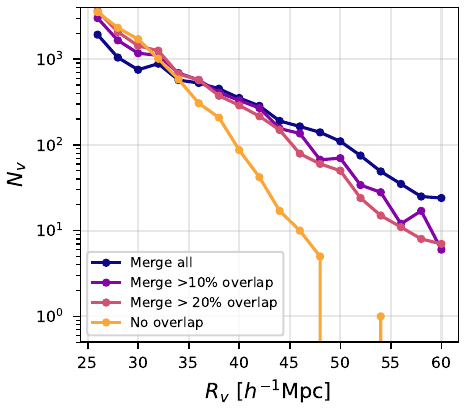}
    \caption{Effect of the void merging fraction on the measured \gls{VSF}. Reducing the fraction of overlap required for merging leads to an increase in the absorption of small voids by larger voids, shifting the void population toward larger radii. The two extremes---prohibiting all overlap or merging every overlapping void---are both unphysical; a more realistic description lies between these limits and can be calibrated using simulations.}
    \label{fig:merging}
\end{figure}

Merging is permitted with any previously detected void and can therefore occur between voids in the same radial bin or in different bins. The parent void is defined as the earliest void in the merger tree and is therefore the most underdense. By construction, child voids are detected later and have higher enclosed densities.
This criterion allows smaller voids that satisfy the density threshold to be incorporated into larger structures, providing a more accurate estimate of the true void volume, which is generally non-spherical due to the non-linear evolution of the density field. As a simple illustrative example, if non-linear evolution stretches a void along one dimension, enforcing strict sphericity will cause it to appear smaller in volume, leading to its classification in a lower size bin. \cite{Paz2023} provide a detailed discussion of how spherical void-finding algorithms fragment non-spherical underdensities and motivate algorithms that can consolidate spherical regions into single void objects. In the case of \code{VERSUS}, allowing merging recovers the true void volume, enabling smaller spheres to fill in the remaining underdensity. Additionally, \citet{Kreisch2022} demonstrated that relaxing the assumption of strict sphericity enhances the information content extracted from void sizes. Moreover, cosmological information from \gls{RSD} and \gls{AP} distortions in observational data can only be properly captured when non-spherical volumes are permitted. 

The impact of void merging on the \gls{VSF} is illustrated in \autoref{fig:merging}. The extreme case of 100\% merging is no more physical than enforcing no merging at all, as genuinely distinct voids, separated by thin filamentary structures of the cosmic web, are expected to exhibit some degree of overlap when their boundary is defined by a sphere. In \autoref{sec:results_abacus}, we find that a 10\% overlap threshold yields a good match to the simulated data, although its applicability to other cosmologies and galaxy-halo parameter choices remains to be tested.

Once all cells satisfying $\delta^\mathrm{sm}<\Delta_v$ have been classified, the initial unsmoothed field is then smoothed again using a \gls{STH} at the next largest input radius. Thus voids are identified in discrete size bins, defined by the user-specified set of input radii. After this process has been repeated for all input radii, the void positions are finalised and voids that have undergone mergers have their radii estimated from the total merged volume $V$, obtained by summing the volumes of the parent void and all associated child voids:
\begin{equation}
    R_v = \left( \frac{3V}{4\pi} \right)^{1/3}.
\end{equation}

The resulting catalogue is consistent with excursion set predictions: void centres are identified from smoothed regions that fall below a specified integrated density contrast, while their volume is allowed to expand to encompass the full extent of any non-spherical underdensities. This accommodates departures from sphericity and recovers the void sizes expected under the spherical, isolated assumptions of \cref{eq:VSF}.

\subsection{Usage}
\label{sec:usage}

\code{VERSUS} has been developed with both performance and usability in mind. The code is open source and can be installed via \code{pip}. The \code{Python} example in \autoref{lst:usage} demonstrates the concise, modular commands through which \code{VERSUS} can be operated; the package also provides a command-line interface, enabling use directly from the terminal. 
\begin{lstlisting}[language=Python, label={lst:usage}, caption={Running \code{VERSUS} on survey data. For use on cubic simulations, simply omit the random catalogue.}]
import numpy as np
from VERSUS import SphericalVoids

# load position catalogues
data = np.load("data.npy")
randoms = np.load("randoms.npy")  # if survey data

# instantiate void-finder
vf = SphericalVoids(data_positions=data,
                    random_positions=randoms,
                    cellsize=4)

# find voids
vf.run_voidfinding(np.arange(25, 62, 2),  # radii
                   void_delta=-0.8,
                   void_overlap=True,
                   void_merge=0.9)

# output
vf.position         # void positions
vf.radius           # void radii
vf.size_function    # VSF
vf.cell_membership  # void membership

\end{lstlisting}
In this example, we use an integrated density threshold of $\Delta_v=-0.8$, a cellsize of $4\Mpch$, input radii in the range $R_v =[25,61]\Mpch$ with a bin width $\Delta R_v =2\Mpch$, and a merging threshold of $f_\mathrm{mg}=0.9$ (i.e. $>10\%$ void overlap is required for merging) with overlap permitted up to void centre. 

\begin{figure}
    \centering
    \includegraphics[width=\linewidth]{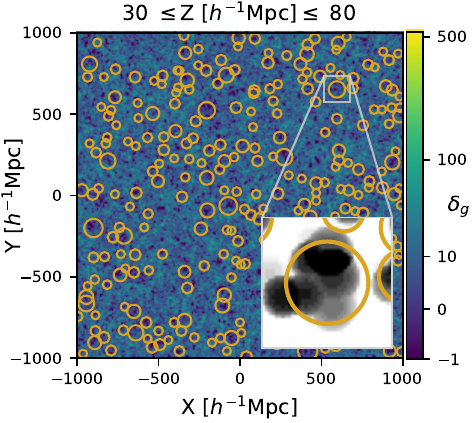}
    \caption{Void distribution in a $50 \Mpch$ thick slice of the \Abacus simulation. The galaxy overdensity field, $\delta_g$, is shown in the background, with darker regions indicating the lowest-density areas. The circles represent the effective spherical radii, rescaled to preserve the merged void volume, although individual voids are in general not spherical. The inset highlights a representative void, with void cells marked in black, explicitly constructed as the union of its constituent overlapping spheres.}
    \label{fig:slice}
\end{figure}

Applying this to the \Abacus simulation described in \autoref{sec:sims_abacus}, \autoref{fig:slice} shows the distribution of voids in a $(2\Gpch)^2$ slice with a width of $50 \Mpch$. The galaxy overdensity field is shown in the background, with darker regions indicating the lowest-density areas. Void centres and sizes are represented by the positions and radii of the circles, tracing the most underdense regions of the galaxy distribution on which they are overlaid. Varying degrees of overlap are evident, with many voids arising from merged ensembles of spheres. The inset highlights one such structure, composed of multiple underdense spheres, where cells associated with voids are marked in black and surrounding unassigned cells are marked in white.

\subsection{Uncertainty in the estimation of void sizes}
\label{sec:bias}

Void-finding algorithms are inherently susceptible to the properties of the tracer field. In particular, the tracer number density can have a significant effect on the sizes of voids detected  \citep{Nadathur2015a}. Therefore, it is important to understand this effect on the \gls{VSF} for a given void-finder in order to perform a robust cosmological analysis. The \gls{VSF} exhibits a steep, quasi-exponential decay towards large radii---posing a problem for finely-binned measurements. Even a symmetric scatter in the estimated radii into neighbouring bins produces a rightward shift in the distribution, as a larger number of voids scatter into a given bin from smaller radii than from larger ones. This is akin to a Malmquist bias \citep{Malmquist1922} in a magnitude-limited sample, where measurement scatter and a steep luminosity function lead to an excess of bright objects due to up-scattering from the more numerous faint population. If left unaccounted for, this problem may bias cosmological inference; however the issue can be alleviated either through accommodating it in the modelling, or by choosing a suitably coarse binning, 

Due to the algorithmic simplicity of \code{VERSUS}, it is possible to write an approximate analytic expression for the uncertainty in the void radius assuming that tracers are randomly spatially distributed around voids. The probability of detecting a void with true radius $R$ using a top-hat smoothing of radius $r$ is dictated by the probability of the integrated density contrast, $\Delta(r)$, being below the void density threshold, $\delta_v$: 
\begin{align}
\begin{split}
        P_v(r\,|\,R) 
        &\equiv P\left[ \Delta(r) \leq \Delta_v \right] ,\\
        &= P\left[ N_t(r) \leq \bar\rho_t(1 + \Delta_v) V(r) \right] .\\
\end{split}
\end{align}
For simplicity, we have rewritten this statement in terms of the enclosed number of tracers $N_t(r)$, the mean tracer density $\bar\rho_t$ and the spherical volume $V(r)=4\pi r^3/3$. Assuming that the enclosed number of tracers can be described by a Poisson process with mean $\bar N_t(r)=\bar\rho_t (1 + \Delta(r)) V(r)$, we can write void detection probability as a Gaussian integral, in the large $\bar N_t(r)$ limit:
\begin{align}
\begin{split}
        P_v(r\,|\,R)
        &= \frac{1}{\sqrt{2\pi \bar N_t}}
        \int_0^{\bar\rho_t(1 + \Delta_v) V}
        \exp\left[{-\frac{(N_t - \bar N_t)^2}{2\bar N_t}} 
        \right] \, dN_t , \\
        &= \frac{1}{2} \left( \mathrm{erf} 
        \left[\frac{\bar\rho_t(1 + \Delta_v) V - \bar N_t}{\sqrt{2\bar N_t}}\right] 
        + \mathrm{erf}\left[\sqrt{\frac{\bar N_t}{2}}\,\,\right]
        \right) , \\
        &\approx \frac{1}{2} \left( \mathrm{erf} 
        \left[\sqrt{\frac{\bar\rho_t V}{2}}
        \frac{\Delta_v - \Delta}{\sqrt{1 + \Delta}}\right] 
        + 1 \right) , \\
        \label{eq:P_void}
\end{split}
\end{align}
where $\mathrm{erf(x)}$ denotes the error function, and we have dropped the $r$-dependence for clarity. 

\autoref{eq:P_void} depends on the integrated void density profile, $\Delta(r)$, which is not known a priori but can be estimated from the data. As such, \code{VERSUS} includes routines for measuring the profile and computing $P_v(r\,|\,R)$. These routines account for potential evolution of the profile with size by evaluating it in bins of $R$. Furthermore, to mitigate potential systematic biases in the estimated void sizes, an internal rescaling is applied to enforce agreement between the integrated density and the threshold value at the void radius. In the case of the synthetic simulations described in \autoref{sec:sims_synthetic}, however, $P_v(r\,|\,R)$ can be computed directly as the true profiles are known. 

Since \code{VERSUS} employs an iterative smoothing procedure, the relevant quantity is the probability that a void is identified in a radius bin $r_i$, given that it has not been detected in any larger bins $r_j$:
\begin{equation}
    P(r_i|R) = P_v(r_i| R) \prod_{j=i+1}^{n_\mathrm{bins}} \left[ 1 -  P_v(r_j| R) \right] .
    \label{eq:P_detect}
\end{equation}
This expression defines a `smearing' function for a void of true radius $R$, representing the detection probability that is assigned to each of the input radius bins. The raw theoretical prediction, discussed in \autoref{sec:vsf_theory}, provides the expected number of voids as a function of radius, $N_v^\mathrm{th}(R)$. By convolving this smearing function with the theoretical prediction, we obtain the expected number of voids measured by \code{VERSUS} in each radius bin:
\begin{align}
\begin{split}
     N_v(r_i |R) 
     &= \int_0^\infty P(r_i|R) \frac{dN_v^\mathrm{th}}{dR} \, dR , \\
     &= \int_0^\infty P(r_i|R) \frac{V}{R}\frac{dn_v}{d\ln{R}} \, dR .
\end{split}
\end{align}

This result provides a powerful framework for quantifying the impact of binning and tracer density on the measured \gls{VSF}, and could in principle be incorporated into the likelihood to enable a more robust analysis. However, it has important limitations beyond the fact that Poisson counts may be a poor approximation for real data: at very low tracer densities, void centres are poorly constrained, introducing additional bias in the inferred radii, and at scales where overlapping voids dominate or the enclosed tracer counts are low, the underlying assumptions of the expression begin to break down. Nonetheless, it is a critical tool to understand the behaviour of \code{VERSUS}.

\subsection{Comparison to other algorithms}

In \autoref{sec:results}, we make direct comparisons to two topological void-finding algorithms: \code{VIDE} \citep{Sutter2015} and \code{REVOLVER} \citep{Nadathur2019c, REVOLVER_2019}.\footnote{\code{REVOLVER} provides two distinct void-finding algorithms: a particle–mesh approach, termed `Voxel', and a Voronoi tessellation–based method, referred to as `Zobov'. In this work, we consider only the latter.} In the context of use on simulation boxes with periodic boundary conditions and a uniform selection function, these codes are essentially algorithmically equivalent, with both making use of the Voronoi-tessellation watershed technique of ZOBOV \citep{Neyrinck2008}, and differ only in their definition of void centre.\footnote{When applied to real survey data, \code{VIDE} and \code{REVOLVER} use somewhat different prescriptions for handling the distorting effects of survey boundaries and selection functions on the density field estimated from the Voronoi tessellation.}

\section{Simulations}
\label{sec:simulations}

To validate the performance of \code{VERSUS}, two types of simulation are considered: (i) a toy scenario consisting of known input voids embedded as underdensities within an otherwise uniform distribution of tracer particles, and (ii) a full $N$-body simulation with dark matter halos populated by mock galaxies based on a realistic galaxy-halo connection model. We use the former \emph{synthetic} void population to assess the ability of \code{VERSUS} to recover the true void centre positions and radii from a discrete tracer field in the presence of shot noise effects, while the $N$-body mock enables tests of its ability to reproduce theoretical predictions in realistic galaxy samples. We further investigate the impact of imposing a complex angular mask in the latter scenario, demonstrating that the code can be straightforwardly applied to survey data with non-trivial geometric footprints.

\subsection{AbacusSummit HOD mock catalogue}
\label{sec:sims_abacus}

The \Abacus simulation suite \citep{ABACUS2021a, ABACUS2021b} provides high-resolution, $N$-body $(2\Gpch)^3$ cubic volumes across a range of cosmologies and redshifts. Constructed using $6912^3$ dark matter particles, the simulations have been designed with the high accuracy of Stage-IV clustering measurements in mind. The suite provides subsampled dark matter particle catalogues and halo catalogues, identified with the \code{CompaSO} algorithm \citep{COMPASO2022}. In this work, we use the redshift $z=0.5$ snapshot output of the simulation with cosmological parameters based on the mean estimates of the \textit{Planck} 2018 TT,TE,EE+lowE+lensing posterior: $\omega_\mathrm{cdm} = 0.1200$, $\omega_\mathrm{b} = 0.02237$, $\sigma_8 = 0.811355$, $n_s = 0.9649$, $h = 0.6736$, $w_0 = -1$, $w_a = 0$, and a single $0.06$ eV massive neutrino \citep{Planck2018}. 

To map the halo catalogue to a realistic distribution of galaxies, we employ the \code{AbacusHOD} code \citep{ABACUSHOD2022}. The code utilises the \cite{Zheng2005, Zheng2007} five-parameter \glscite{HOD}{Berlind2002} model where the mean occupation numbers of central and satellite galaxies in a halo of mass $M_h$ are given by
\begin{align}
    \bar{n}_{\mathrm{cent}}(M_h) & = \frac{1}{2}\,\mathrm{erfc} \left[\frac{\log_{10}(M_{\mathrm{cut}}/M_h)}{\sqrt{2}\sigma}\right] , \label{eq:zheng_hod_cent}\\
    \bar{n}_{\mathrm{sat}}(M_h) & = \left[\frac{M_h-\kappa M_{\mathrm{cut}}}{M_1}\right]^{\alpha}\bar{n}_{\mathrm{cent}}^{\mathrm{LRG}}(M_h) .
    \label{eq:zheng_hod_sat}
\end{align}
Here, the conditional error function $\mathrm{erfc}(x)\equiv 1-\mathrm{erf}(x)$. Mass thresholds $M_{\mathrm{cut}}$ and $\kappa M_\mathrm{cut}$ set the minimum halo mass to host a central galaxy and satellite galaxy, respectively. $M_1$ is approximately the typical halo mass that hosts a single satellite galaxy. The transition from empty to central-hosting halos is dictated by the value of $\sigma$ while the exponent $\alpha$ controls the slope of the satellite occupation distribution. 

Using this prescription, we populate the halo catalogue with a model that provides an excellent fit to early \gls{DESI} measurements \citep{Yuan2024}. The parameter values used in this work were taken from the posterior mean of a fit to the projected clustering of the \gls{LRG} sample between $0.4<z<0.6$ in the \gls{DESI} One-Percent Survey \citep{DESI2024} and are summarised in Table 3 of \cite{Yuan2024}: $\log M_\mathrm{cut}=12.89$, $\log M_1=14.08$, $\sigma=0.27$, $\alpha=1.20$, and $\kappa=0.65$. We leave explorations of other \gls{HOD} models to further work. Finally, the galaxy population is subsampled to approximately match the DESI \gls{LRG} number density, $n \approx 5\times 10^{-4}\hhhMpc$.

\subsection{Synthetic void population}
\label{sec:sims_synthetic}

\begin{figure}
    \centering
    \includegraphics[width=\linewidth]{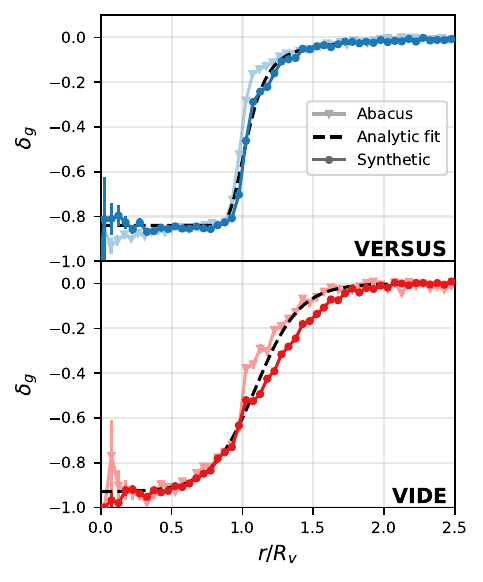}
    \caption{Void density profiles scaled by the void radius as measured by \code{VERSUS} (\textit{blue}) and \code{VIDE} after applying the cleaning algorithm (\textit{red}) using the simulations discussed in \autoref{sec:simulations}. The analytic curves (\textit{dashed black}), given by \autoref{eq:profiles}, correspond to symbolic regression fits to profiles measured from the \Abacus simulations (\textit{faint triangles}). This analytic profile is then used as input for the synthetic void simulations, with the resulting recovered density profile shown by the \textit{circular} markers.}
    \label{fig:profiles}
\end{figure}

As discussed in \autoref{sec:bias}, the distribution of measured voids is naturally sensitive to the number density of tracers from which they are determined. Depending on the void-finding algorithm, this may present in different ways. At minimum, it adds a degree of stochasticity into the estimation of the void radius---particularly critical for studies of the \gls{VSF} where noise in the binned measurement can lead to a skewing of its quasi-exponential shape. To ensure that \code{VERSUS} correctly identifies the properties of spherical underdensities predicted by the excursion set theory, we generate a set of random tracer distributions containing synthetic void populations.

To investigate the impact of sample number density, we populate $(2 \Gpch)^3$ cubic volumes with a uniform distribution of tracers, exploring a range of number densities around $n \sim 5 \times 10^{-4} \hhhMpc$. A set of candidate void centres is then randomly selected within this volume. Using 100 radial bins, we sample the theoretical \gls{VSF} in the range $R_v =[35, 55]\Mpch$ to determine the expected number of voids in each bin. We evaluate the theory prediction with $b_1=2.1$, consistent with the \gls{DESI} \gls{LRG} tracer, and $B_\mathrm{slope}=0.96$ and $B_\mathrm{offset}=0.26$, matching the values reported in \cite{Contarini2022}. This distribution is subsequently used to assign radii to the candidate centres, subject to the condition that newly assigned voids did not overlap with those previously placed. Following the assignment of a radius $R$, we remove tracer particles within a spherical region of radius $2.5R$ and replaced them with particles sampled from a prescribed density profile. 

The density profiles used to embed depressions in the random distribution are calibrated to simulations. We measure profiles using the \Abacus simulation, described in \autoref{sec:sims_abacus}, identifying voids either directly with the \code{VERSUS} algorithm or from the cleaned \code{VIDE} catalogue. Analytic forms of these profiles are obtained via symbolic regression using the \code{pyoperon}\footnote{\url{https://github.com/heal-research/pyoperon}} algorithm, applied to stacked void profiles in the range $R_v =[35,40]\Mpch$. We choose to use symbolic regression to obtain analytic forms approximately reproducing the measured profiles in each case in order to make the results easier to replicate in other studies. As the general lessons drawn from the exercise are not sensitive to the precise details of the void density profile, we also prioritised obtaining a relatively simple closed-form expression for the density profiles over exactly matching the \Abacus results. Both \code{VERSUS} and \code{VIDE} post-cleaning should find voids that enclose the chosen density threshold. Therefore, we perform a minor adjustment to the profiles to ensure that the integrated results intersect $\Delta(r=R)=-0.8$. We determine the following analytic profiles for \code{VERSUS} and cleaned \code{VIDE} voids:
\begin{align}
    \delta_\mathrm{versus}(r_\mathrm{s}\equiv r/R)  &=
    0.84 \left[ \exp\left(
    -5.5\times 10^{4} e^{-11 r_s} - e^{-2 r_s} 
    \right) - 1 \right] , \nonumber \\
    \delta_\mathrm{vide}(r_\mathrm{s}\equiv r/R)  &=
    0.93 \left[ \big(
    1 + \exp(7 - 6.4 r_s)
    \big)^{-1} - 1 \right] .
    \label{eq:profiles}
\end{align}
Note that here we use the notation $\delta(r)$ to refer to the angle-average density contrast in a spherical shell of radius $r$ from the void centre, which is distinct from the integrated density contrast $\Delta(r)$ within a sphere of radius $r$. Before embedding the profile, we rescale them by the radius assigned to each candidate void.

\begin{figure*}
    \centering
    \includegraphics[width=\linewidth]{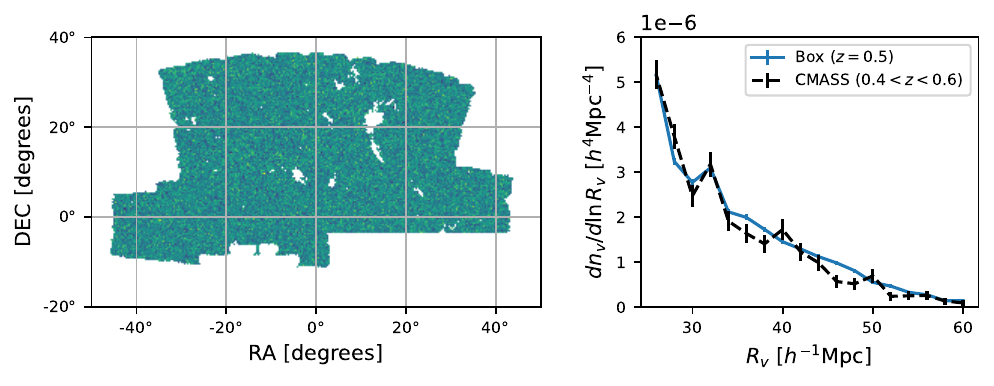}
    \caption{Effect of the survey footprint on the measured \gls{VSF}. \textit{Left}: angular distribution of the random catalogue constructed from the \gls{BOSS} DR12 CMASS \gls{SGC} mask. The same angular mask and redshift-dependent $n(z)$ selection are applied to both the data and random catalogues. \textit{Right}: \gls{VSF} measurements from the cubic \Abacus mock downsampled to match the median $n(z)$ (\textit{blue}) and the \Abacus mock after imposing the survey footprint (\textit{black dashed}).}
    \label{fig:cutsky}
\end{figure*}

We generate simulations for both \code{VERSUS} and \code{VIDE} profiles with a total number of $486$ voids at different tracer number densities. \autoref{fig:profiles} shows the analytic profiles compared to the \Abacus void profiles on which they are based. Also plotted are the profiles measured from the synthetic void population with a tracer density $n\approx5\times 10^{-4} \hhhMpc$, equivalent to that of the \gls{DESI} \gls{LRG} sample. If the chosen void-finding algorithm correctly estimates the void centres and radii, then the measured result will exactly match the analytic profile.

\subsection{Mocks with a complex survey footprint}
\label{sec:sims_cutsky}

For simplicity, the majority of the performance tests of \code{VERSUS} reported in this work are based on its application to simulated data in cubic boxes with periodic boundary conditions and a uniform selection function. However, the algorithm is designed such that it can readily be applied real survey data with a complex survey geometry and selection function, through the use of an additional unclustered `random' catalogue which incorporates these non-cosmological effects as commonly produced for large-scale structure analyses. 

To validate the performance of the algorithm in such scenarios, we cut out a section of the cubic \Abacus mock described in \autoref{sec:sims_abacus} to create a survey-like mock. This mock is chosen to have an angular mask constructed to match that of the  \gls{BOSS} DR12 CMASS \gls{SGC} galaxy sample, with an irregular geometry and survey holes, as shown in \autoref{fig:cutsky}. We translate the mock galaxy positions relative to an observer placed close to the edge of the box into redshifts, and then cut the mock to include only redshifts in the range $0.4 < z < 0.6$. We also impose a redshift-dependent selection function by randomly downsampling the mock galaxies in order to match the radial number density, $n(z)$, of the CMASS data. 

To construct a matching random catalogue incorporating these selection effects, we repeat the same cuts after starting with an unclustered set of random points with a number density 50 times larger than that of the mock galaxies. This procedure captures the main complications of void-finding applied to real survey data and is sufficient to demonstrate the performance of \code{VERSUS}. While additional observational effects such as angular variation in the sample completeness could also be included for even greater realism, we do not expect them to materially alter the conclusions drawn below as long as any such effects are accurately captured by the associated random catalogue, if necessary by the inclusion of systematic weights. The accuracy of the random catalogue is a condition for \emph{any} reliable large-scale structure measurements from galaxy redshift surveys and is not specific to void-finding with \code{VERSUS}, so this is a rather minimal assumption.

\section{Results}
\label{sec:results}

In this section, we present the results of the \code{VERSUS} validation tests on simulated data, described above in \autoref{sec:simulations}, assessing its ability to accurately recover void sizes and reproduce the theoretically predicted \gls{VSF}. We also examine how performance and computational efficiency vary with different input settings.

In the following tests, we use \code{VERSUS} with a cellsize of $4\Mpch$ ($N_\mathrm{grid}=500^3$), input radii binned in the range $R_v =[25,61]\Mpch$ with a width $\Delta R_v =2\Mpch$, and a merging threshold of $f_\mathrm{mg}=0.9$, unless otherwise specified.

\subsection{Performance and validation}
\label{sec:results_performance}

\autoref{fig:cutsky} illustrates the ability of \code{VERSUS} to handle the complex survey footprint of the \gls{BOSS} DR12 CMASS \gls{SGC} sample. As the \gls{VSF} is sensitive to the tracer number density, we apply the survey geometry and redshift-dependent $n(z)$ selection to the \Abacus mock, and compare to the cubic volume case with a downsampling factor of $1.8$ to match the median $n(z)$. By using a catalogue of random positions that trace this geometry, generated at 50 times the density of the galaxy sample, \code{VERSUS} accurately recovers the \gls{VSF} measured from the downsampled cubic mock. This result firmly establishes the readiness of the code for robust application to real survey data. 

\begin{figure}
    \centering
    \includegraphics[width=\linewidth]{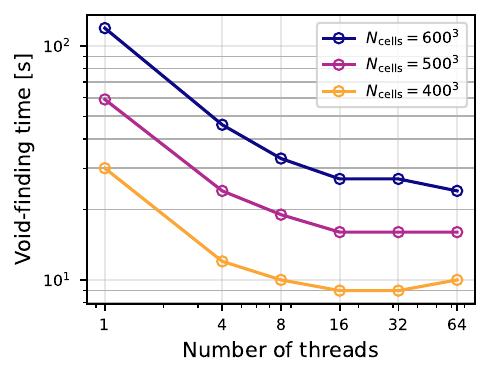}
    \caption{Wall time for the void-finding step of \code{VERSUS} as a function of the number of multi-threaded processes. The measurements correspond to the setup described at the start of \autoref{sec:results}, with variations in the number of grid cells $N_\mathrm{cells}$ illustrated. In this work, we adopt $N_\mathrm{cells}=500^3$, corresponding to a cell resolution of $4\Mpch$ on the $(2\Gpch)^3$ simulations.}
    \label{fig:speed}
\end{figure}

\autoref{fig:speed} shows the efficiency of the \code{VERSUS} algorithm for different grid resolutions as a function of the number of multi-threaded processes. The wall times shown correspond to the measurements for the void-finding step alone (i.e. after mesh initialisation) and therefore provide an estimate of the runtime required for different void-finding configurations (e.g. void density threshold, input radii, merging threshold, etc.). All runs were performed on a single AMD EPYC 7763 CPU node on the NERSC Perlmutter cluster\footnote{\url{https://docs.nersc.gov/systems/perlmutter/architecture/}} which has 128 cores. By default, \code{VERSUS} will use all available CPU cores using \code{OpenMP} parallelization.

Setting $N_\mathrm{threads}=16$ is optimal for achieving evaluation times under 30 seconds for all the grid sizes considered. In this work, we have demonstrated that the desired accuracy on $(2 \Gpch)^3$ simulations can be achieved with $N_\mathrm{cells}=500^3$, corresponding to runtimes below 20 seconds. However, a modest reduction in resolution\footnote{Note that while the runtime time scaling is $\mathcal{O}(N_\mathrm{cells})$, the resolution scales only as $N_\mathrm{cells}^{1/3}$.} can further reduce runtimes to just a few seconds---enabling the full potential of simulation-based methods to be exploited. For an input mock with $\sim4\times10^6$ galaxies, using $N_\mathrm{threads}=8$, $N_\mathrm{cells}=500^3$ and running on 8 cores of an Intel Core i7-1185G7 CPU, \code{VERSUS} achieves a gain of $\sim8\times$ in speed relative to \code{VIDE}. While computation times for \code{VIDE} scale roughly proportional to $N_\mathrm{gal}$,\footnote{Nico Schuster, private communication.} for \code{VERSUS} the corresponding scaling is $\propto N_\mathrm{cells}$.

\subsection{Synthetic void population}
\label{sec:results_synthetic}

\begin{figure}
    \centering
    \includegraphics[width=\linewidth]{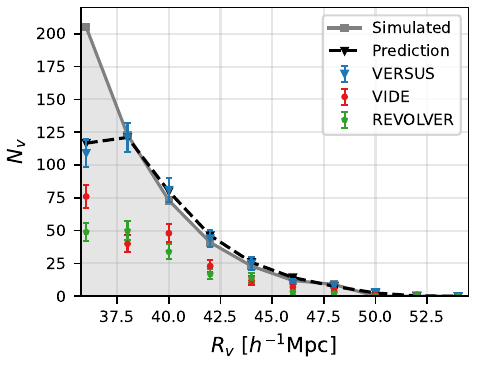}
    \caption{\gls{VSF} measurements obtained with different algorithms using the synthetic simulations described in \autoref{sec:sims_synthetic}, which contain $486$ voids in a random distribution of tracers with number density $n \approx 5\times 10^{-4} \hhhMpc$. Results for each algorithm are shown as coloured points. Both \code{VIDE} and \code{REVOLVER} were applied to the same tracer distribution and post-processed following the cleaning procedure of \protect\cite{Ronconi2017}. The input synthetic distribution is shown in \textit{grey}, while the \textit{dashed black} line indicates the analytic prediction for \code{VERSUS} as described in \autoref{sec:bias}. Error bars denote the Poisson uncertainty on the void number counts.}
    \label{fig:synthetic_fit}
\end{figure}

\begin{figure*}
    \centering
    \includegraphics[width=\linewidth]{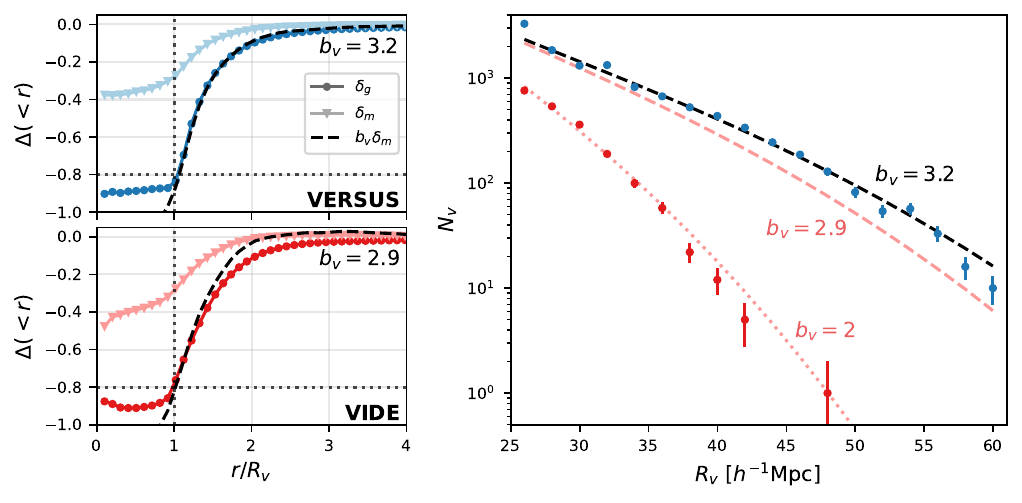}
    \caption{\textit{Left}: Integrated galaxy (\textit{circles}) and matter (\textit{faint triangles}) density profiles for voids measured in the galaxy distribution using the \code{VERSUS} (\textit{blue}) and cleaned \code{VIDE} (\textit{red}) algorithms. The biased matter profiles (\textit{dashed black}) using $b_v=3.2$ and $b_v=2.9$, for \code{VERSUS} and \code{VIDE}, respectively, provide a good fit to the galaxy profiles at $\Delta(r=R) = \Delta_v$, marked by the \textit{dotted} crosshairs. 
    \textit{Right}: The corresponding \gls{VSF} measurements for \code{VERSUS} (\textit{blue}) and \code{VIDE} (\textit{red}). The \code{VERSUS} measurements are well described by a theoretical prediction with $b_v=3.2$ (\textit{dashed black}). The cleaned \code{VIDE} measurement only matches the theory prediction when setting $b_v\sim2$, inconsistent with the true value of $b_v=2.9$ determined from the density profile. Error bars denote the Poisson uncertainty on the void number counts.}
    \label{fig:abacus_fit}
\end{figure*}

\autoref{fig:synthetic_fit} shows the resulting void size distributions from \code{VERSUS}, \code{VIDE}, and a similar topological algorithm, \code{REVOLVER}, applied to the synthetic void population. The \code{VIDE} and \code{REVOLVER} distributions have been cleaned using the method of \cite{Contarini2023} intended to improve the match to the excursion set predictions, which reduces the total number of voids to approximately $0.5\%$ of the raw catalogue. The true \gls{VSF} of the simulated synthetic population, described in \autoref{sec:sims_synthetic}, is displayed as the filled grey curve. The analytic prediction for \code{VERSUS}, given the binning and tracer number density, was calculated using the method described in \autoref{sec:bias} and is displayed by the black dashed line. 

As expected, the \code{VERSUS} \gls{VSF} is excellently described by the analytic prediction in the case of randomly distributed tracer particles. It is also in very good agreement with the simulated truth across all but the smallest radius bin. This apparent paucity of voids at $R\sim 36\Mpch$ is a direct result of uncertainty in the radius determination, which scatters a fraction of the measured radii into neighbouring bins. As no voids with $R<35\Mpch$ were simulated, this scattering is unidirectional. While this unphysical cutoff in radii does not affect more realistic scenarios, it is nonetheless successfully captured by the analytic prediction.

Given the \gls{DESI} \gls{LRG}-like number density of the simulation, we find that a bin width of $\Delta R_v = 2\Mpch$ is sufficient to recover the simulated truth with \code{VERSUS}.
At lower number densities ($n\sim 1\times10^{-4} \hhhMpc$), the uncertainty in the radius increases, producing a rightward shift in the distribution for fixed binning. Owing to the quasi-exponential form of the \gls{VSF}, a symmetric scattering into neighbouring bins contributes a larger fractional excess to bins at higher radii. However, $\Delta R_v = 2\Mpch$ is sufficiently coarse to suppress this skewness for $n\approx5\times 10^{-4} \hhhMpc$, such that the analytic correction can be neglected. It remains to be seen if this holds for cosmologies with a much steeper \gls{VSF}.

The cleaning procedure applied to the raw \code{VIDE} and \code{REVOLVER} catalogues consists of three steps: (i) removing voids with high central densities to eliminate those identified from spurious density fluctuations, (ii) rescaling void radii so that they enclose the chosen density threshold based on their density profile measured from the void centre, and (iii) resolving overlaps by discarding the void with the higher central density. Applied to \code{VIDE}, the first step recovers $443$ voids---close to the true number of input voids. However, since the excursion set formalism does not distinguish between voids based on their interior densities (i.e. cloud-in-void modes are permitted), this central density criterion over-cleans the sample, discarding $43$ genuine voids. Additionally, a further 96 voids are excluded because noise in the profile determination causes the rescaling to fail.

We observe similar behaviour for \code{REVOLVER}---in this case only differing from \code{VIDE} in the definition of the void centre used in the cleaning procedure for density estimation---when applied to the same simulation as \code{VIDE}. Given the agreement between the \code{VIDE} and \code{REVOLVER} results, we infer that their discrepancy with the truth is primarily driven by the aggressive removal of genuine voids by the cleaning algorithm, rather than by any misalignment between the measured void centres and the true density minima. However, the residual effect of the latter `miscentering' is still evident in \autoref{fig:profiles}: the \code{VERSUS} profile is accurately recovered, whereas the profile of the cleaned \code{VIDE} voids approaches the mean density more gradually at distances larger than the void radius $R_v$ when compared to the analytic input profile. Averaging over many miscentered voids suppresses the underlying profile recovered by \code{VIDE}. Additionally, a characteristic bump at $r=R_v$ in the \code{VIDE} profile, which is evident in both the synthetic and \Abacus simulation measurements, arises when a single tracer causes the density threshold to be exceeded, setting the void radius as it is expanded during the cleaning procedure.

\subsection{AbacusSummit HOD mock catalogue}
\label{sec:results_abacus}

We applied \code{VERSUS} to the \Abacus \gls{LRG}-like cubic simulation described in \autoref{sec:sims_abacus}. \autoref{fig:abacus_fit} highlights the performance of \code{VERSUS} in extracting a void sample that is in direct agreement with theoretical predictions. 

The left-hand panel of \autoref{fig:abacus_fit} shows the measured integrated density profiles for each algorithm. Measured around the same void centres, the corresponding dark matter profiles are also displayed. Following the approach of \cite{Contarini2019}, evaluating the tracer bias at the void radius provides a means to translate the theoretical prediction from the matter field to the tracer field (see \autoref{eq:bias}). We find that similar void bias values of $b_v=3.2$ and $b_v=2.9$, for \code{VERSUS} and \code{VIDE}, respectively, provide good fits to the galaxy density profiles near the void radius.

The right-hand panel shows the \gls{VSF} as measured by both \code{VERSUS} and \code{VIDE} post-cleaning. There is a large disparity between the two distributions, qualitatively similar to that observed in \autoref{fig:synthetic_fit}. The dashed and dotted lines show theoretical predictions, computed at the true cosmology of the simulation following \autoref{eq:VSF}, but with different void bias values (\autoref{eq:bias}) as indicated in the figure. Both measured \glspl{VSF} can therefore be accommodated within the same cosmological model, albeit with different bias values. While the \gls{VSF} measured by \code{VERSUS} is self-consistently fit by the theoretical model using the value of $b_v=3.2$ implied by its density profile, the \code{VIDE} \gls{VSF} favours a significantly lower bias ($b_v\sim 2$)---inconsistent with the value of $b_v=2.9$ inferred from its density profile.

Beyond consistency of the theoretical framework, \autoref{fig:abacus_fit} also demonstrates that \code{VERSUS} results in an improved signal-to-noise measurement since it provides a much more numerous catalogue of voids from the same galaxy sample, thus decreasing shot noise in the VSF. However, we leave a full Bayesian analysis to future work.

\section{Conclusions}
\label{sec:conclusions}

This work presents \code{VERSUS}, a novel void-finding algorithm designed to identify spherical underdensities that directly correspond to excursion set predictions of the \gls{VSF}. The method is fast, user-friendly, and conceptually simple. By applying a series of iterative top-hat smoothings to the density field, \code{VERSUS} generates a population of voids that are well described within the excursion set framework. When combined with a straightforward merging criterion, the algorithm can recover the effective radii of voids that have become distorted from their initial spherical shapes through non-linear evolution. The merging procedure is conceptually similar to the ideas presented in \cite{Paz2023} and is designed to mitigate the fragmentation of non-spherical underdensities inherent to strictly spherical void-finding algorithms. Owing to its simplicity, an analytic expression for the uncertainty in radius estimates as a function of tracer number density was derived in \autoref{sec:bias}.

By applying \code{VERSUS} to different simulations, we have demonstrated that the resulting void population is well described by the theoretical model. Tests using randomly distributed tracer particles with an imposed synthetic void population show that \code{VERSUS} accurately recovers both void positions and radii in the presence of a discrete tracer field. Notably, this relatively simple toy model result is not easily reproducible with current topological algorithms.  Extending to a more realistic scenario, \code{VERSUS} was applied to an \Abacus dark matter simulation populated with a galaxy-halo connection calibrated to the \gls{DESI} One-Percent Survey. In this setting, the algorithm performs strongly, achieving an excellent match to the theoretical \gls{VSF} without the need for unphysical adjustments of the bias parameter. 

Furthermore, \code{VERSUS} successfully accounted for the non-trivial \gls{BOSS} CMASS DR12 \gls{SGC} survey geometry and the redshift-dependent selection, accurately recovering the \gls{VSF} by employing a catalogue of random points to characterise the survey footprint---a promising step towards an application to real data. This is crucially important for both analytic methods, as it removes the need for ad hoc corrections to the theoretical prediction, and for simulation-based approaches, where a large number of simulations are required and avoiding the extra time-consuming step of creating mocks matching the complex survey footprint from simulations originally generated in a cubic box can be highly advantageous.

This work does not consider observational effects on the void catalogue such as \gls{RSD} and \gls{AP} distortions. These effects are expected to be included within the theoretical modelling in a manner similar to previous works (e.g. \citealt{Correa2021, Contarini2022, Verza2025}), since distortions to the void volume can be captured by the algorithm through its merging criterion. However, further investigation into the impact of a fixed merging parameter across simulations with alternative cosmologies and varied galaxy-halo prescriptions is required. In addition, the code employs the \code{pyrecon} density field reconstruction algorithm, which is expected to improve the recovery of real-space void centres from redshift-space data \citep{Nadathur2019a, Nadathur2019c}---critical for accurate radius estimates. A detailed investigation of observational effects, as well as variations in cosmological and galaxy-halo parameters, is deferred to future work.

The ability of \code{VERSUS} to reproduce theoretical predictions of the \gls{VSF} has been demonstrated in detail throughout this work, across a range of controlled and realistic scenarios. This agreement highlights both the robustness of the algorithm and its suitability for precision cosmological applications. Looking ahead, the computational efficiency of \code{VERSUS} opens the door to more advanced inference techniques. In a follow-up paper, \cite{Findlay2026prep}, this speed will be exploited to train a simulation-based model for void statistics within a fully Bayesian framework, enabling a robust exploration of a broad range of cosmological and galaxy-halo parameters. The distinctive capability of \code{VERSUS} to rapidly generate \gls{VSF} measurements that align directly with theoretical predictions establishes a novel pathway for unifying analytic and simulation-based approaches, enhancing the robustness of cosmic void cosmology.

\section*{Acknowledgements}

NF acknowledges support from STFC grant ST/X508688/1 and funding from
the University of Portsmouth. NF would like to thank Harry Desmond for guidance on performing symbolic regression, Nico Schuster for assistance with \code{VIDE}, Nathan Cruickshank for helping to devise the \code{VERSUS} acronym, and Hernan Rincon for useful discussions.

\section*{Data Availability}

The algorithms used in this paper are available from the \code{VERSUS} repository (\url{https://github.com/ntbfin00/VERSUS}). Figure data and plotting routines are provided in a separate repository (\url{https://github.com/ntbfin00/Paper_figures}). Additional data can also be obtained from the lead author upon request.


\bibliographystyle{mnras}
\bibliography{references}





\bsp	
\label{lastpage}
\end{document}
